\documentstyle{memsait}
\def \SAIT #1 #2 {{\em Mem.\ Soc.\ Astron.\ It.\/} {\bf #1}, #2}
\def \MESS #1 #2 {{\em The Messenger\/} {\bf #1}, #2}
\def \ASTRNACH #1 #2 {{\em Astron. Nach.\/} {\bf #1}, #2}
\def \AAP #1 #2 {{\em Astron. Astrophys.\/} {\bf #1}, #2}
\def \AAL #1 #2 {{\em Astron. Astrophys. Lett.\/} {\bf #1}, L#2}
\def \AAR #1 #2 {{\em Astron. Astrophys. Rev.\/} {\bf #1}, #2}
\def \AAS #1 #2 {{\em Astron. Astrophys. Suppl. Ser.\/} {\bf #1}, #2}
\def \AJ #1 #2 {{\em Astron. J.\/} {\bf #1}, #2}
\def \ANNREV #1 #2 {{\em Ann. Rev. Astron. Astrophys.\/} {\bf #1}, #2}
\def \APJ #1 #2 {{\em Astrophys. J.\/} {\bf #1}, #2}
\def \APJL #1 #2 {{\em Astrophys. J. Lett.\/} {\bf #1}, L#2}
\def \APJS #1 #2 {{\em Astrophys. J. Suppl.\/} {\bf #1}, #2}
\def \APSS #1 #2 {{\em Astrophys. Space Sci.\/} {\bf #1}, #2}
\def \ASR #1 #2 {{\em Adv. Space Res.\/} {\bf #1}, #2}
\def \BAIC #1 #2 {{\em Bull. Astron. Inst. Czechosl.\/} {\bf #1}, #2}
\def \JSQRT #1 #2 {{\em J. Quant. Spectrosc. Radiat. Transfer\/} {\bf #1}, #2}
\def \MN #1 #2 {{\em Mon. Not. R. Astr. Soc.\/} {\bf #1}, #2}
\def \MEM #1 #2 {{\em Mem. R. Astr. Soc.\/} {\bf #1}, #2}
\def \PLR #1 #2 {{\em Phys. Lett. Rev.\/} {\bf #1}, #2}
\def \PASJ #1 #2 {{\em Publ. Astron. Soc. Japan\/} {\bf #1}, #2}
\def \PASP #1 #2 {{\em Publ. Astr. Soc. Pacific\/} {\bf #1}, #2}
\def \NAT #1 #2 {{\em Nature\/} {\bf #1}, #2}
\def \NEWA #1 #2 {{\em New Astron. \/} {\bf #1}, #2}
\input epsf.sty
\begin{opening}
\title{CLUMPUSCULE FORMATION AT HIGH REDSHIFT}
\author{F. Combes}
\institute{Observatoire de Paris, DEMIRM\\
           61 Av. de l'Observatoire, F--75\,014 Paris, France}
\author{D. Pfenniger}
\institute{Observatoire et Universit\'e de Gen\`eve\\
          CH--1290 Sauverny, Switzerland}
\date{} % DO NOT INSERT ANY DATE HERE !!!
\end{opening}

\begin{document}

%\oddpagefooter{\sf Mem. S.A.It., Vol. ??, ??}{}{\thepage}
%\evenpagefooter{\thepage}{}{\sf Mem. S.A.It., Vol. ??, ??}
\oddpagefooter{}{}{} % LEAVE AS IT IS !
\evenpagefooter{}{}{} % LEAVE AS IT IS !
\ 
%\bigskip

\begin{abstract}
Over the last two decades, realistic studies have often conclude 
that the first bound objects to form can be very small, much smaller
than a solar mass.  After recombination, the Jeans mass 
drops rapidly to 
the order of a Giant Molecular Cloud (GMC) mass ($\sim
10^5$\,M$_\odot$), and the H$_2$ cooling can make the collapse
quasi-isothermal;
this leads to recursive fragmentation, and formation of clumps so
dense that 3-body reactions transform the gas almost
entirely to the molecular phase.  This could lead to star formation in
some places, but since
star formation is very inefficient, 
most of the molecular gas could consist of a fractal built on
clumpuscules thermalized with the background radiation, and filling a
tiny fraction of the volume.
{\em The bulk of the gas mass can therefore be trapped in this phase, well
before the first stars 
re-heat  and re-ionize
the diffuse gaseous medium}. This results in a very contrasted
multi-phase baryonic medium, that has partly remained until the present
time.
\end{abstract}

\section{The first bound structures}
\subsection{General background}
In standard big-bang cosmologies, 
observable structures are assumed to form 
from small density fluctuations, that first expand with the Universe
while growing linearly, then collapse non-linearly to bound systems. These
fluctuations could be adiabatic, in which case they would be damped on scales
below $3\cdot 10^{13} (\Omega h^2)^{-5/4}$\,M$_\odot$ before recombination
(Silk 1968; where $h$ is the Hubble constant in units of 
100\,km\,s$^{-1}$\,Mpc$^{-1}$). 
Isothermal fluctuations can also exist
(and, of course, also intermediate states), and in such cases, 
the first non-linear objects
are much smaller. If the density fluctuations spectrum is assumed
self-similar, with a slope $n$ such that
\begin{equation}
\frac{\delta\rho}{\rho} \propto M^{-(n+3)/6}
\end{equation}
then the largest masses to become non-linear after recombination, i.e.
at $z\sim 1500$, are of the order of $M \sim 10^{6-8}$ M$_\odot$,
depending on $\Omega$, $h$ and $n$, if the spectrum is calibrated from
the presently observed correlation function (e.g., Peebles 1980).  That
means that all masses smaller than $M \sim 10^{6-8}$\,M$_\odot$, and
larger than the Jeans mass, which is $M \sim 10^5$\,M$_\odot
({{\Omega_{\rm b}}\over{0.06}})^{-1/2} ({{h}\over{0.5}})^{-1}$ at $z=1000$,
will collapse, become bound, and decouple from expansion (at this
epoch the Jeans length is of the order of 2\,pc\,$({\Omega_{\rm
b}\over 0.06})^{-1/2} ({h\over 0.5})^{-1})$.  In fact, since in every
hierarchical model with $n>-3$ the smallest masses have more density
contrast, structures at precisely the Jeans mass collapse first.  The
masses correspond to typical $z=0$ giant molecular clouds.

The nature of the collapse has been studied by many authors. If
cooling is efficient enough ($\tau_{\rm cool} \sim \tau_{\rm ff}$),
the collapse is quasi-isothermal, and fragmentation occurs, since the
Jeans mass becomes smaller and smaller
as the density increases (e.g., Hoyle 1953). This means that pressure
forces are at most equal to gravity forces, and cannot stop the
collapse. This fragmentation process occurs currently 
in the Galactic interstellar medium
in a self-similar hierarchical structure (Larson 1981; Scalo 1985),
and might even be pursued down to
very low masses in particularly cold and quiet regions, such as the
outer parts of galaxies (Pfenniger \& Combes 1994). Fragmentation is
limited by opacity, and the smallest fragments (or clumpuscules),
which are at the transition of being 
pressure supported, are today of the order of $10^{-3}$ M$_\odot$, and
their mass grows slowly, as $T^{1/4}$ or $(1+z)^{1/4}$, with redshift.
This assumes a quasi-isothermal regime for
the bottom of the hierarchy (containing most of the mass);
therefore it is important that gas cooling remains efficient.  

This might look as 
a problem for gas just after recombination, since it is metal poor
without grains: 
if above $10^4$\,K the main coolant is atomic hydrogen (by collisional
excitation of Ly$\alpha$), in fact below this temperature the
vibration-rotation lines of molecular hydrogen take over until
$T=200$\,K, because a significant quantity of H$_2$ molecules 
forms through H$^-$ and H$_2^+$.  Below 200\,K, HD is then more
efficient (Palla \& Zinnecker 1987; Puy \& Signore 1996).
Many groups have tackled the problem of computing 
the physico-chemistry of the primordial gas, 
to determine the size of the first forming bound structures (Yoneyama
1972; Hutchins 1976; Carlberg 1981; Palla et al.\ 1983; Lepp \& Shull
1984). All of them have found that the cooling is indeed efficient, as
soon as the redshift is below $z\sim 200$, and the calculated masses
of fragments are in the wide range $0.1 -100$\,M$_\odot$.  Even if
clouds of just the Jeans mass see their collapse somewhat delayed
because of pressure forces, the latter are negligible for clouds of
$M> 10^6$\,M$_\odot$ (Lahav 1986).  In more realistic scenarios, the
collapse is not spherical, but does form sheets and filaments (Larson
1985). The latter have the advantage that in 1D, pressure forces have
always the same dependence with radius than the gravity forces, and
therefore cannot halt the collapse, whatever the cooling (Uehara et
al.\ 1996; Inutsuka \& Miyama 1997). Therefore fragmentation is
inevitable.

The conclusions drawn from these studies were all concerning the first
generation of stars, since {\em no 
outcome to fragmentation other than star formation was considered}.
Already Peebles \& Dicke (1968) concluded that the first objects to
form were the globular clusters, through such a fragmentation process,
followed by star formation.  Palla et al.\ (1983) considered the H$_2$
formation through the very efficient 3-body process, when the density
is higher than $10^8$\,cm$^{-3}$. They concluded that all the
primordial gas is converted in H$_2$ at a density of
$10^{12}$\,cm$^{-3}$, and that the Jeans mass eventually falls below
0.1\,M$_\odot$, allowing the whole mass spectrum for the first forming
stars. Rotation could 
hinder the collapse (Kashlinsky \& Rees 1983), and this could lead to
Population III low-mass stars as well as super-massive objects (VMO's),
both of which could contribute to dark matter.

Taking the observations of the ISM at face, which show that 
the efficiency of star formation is very low, we propose here that
the bulk of the gas in this first collapse of GMC-size primordial
clouds after recombination and decoupling form clumpuscules, assembled
in a hierarchical or fractal structure. The molecular density of these
clumpuscules should be very high ($10^{10}$ to $10^{12}$\,cm$^{-3}$),  
when at the limit of being
pressure supported, at the temperature of the cosmic background.

\subsection{Non star-formation}
Although theoretically not well understood, observations show that the
efficiency of star formation is highly variable, and must be very low
in some cases (e.g., in low surface brightness galaxies such as Malin
1). Since the physical conditions of the gas at $z \sim 50-100$
is similar to the present outer parts of galaxies, we expect that the
conditions for triggering global star formation are generally
not met.  Larger scales structures such as the galaxy-size ones, have
not then collapsed and decoupled from expansion, therefore there does
not exist a deep enough potential well to accumulate mass to the critical
density for star-formation (Kennicutt 1989).  Indeed, star formation
is hindered when the medium is only mildly unstable, 
so that pressure forces compensate gravity forces. That is the case,
when the collapsing masses are always of the order of the Jeans
mass. In the first collapse of the giant molecular clouds of
$10^6$\,M$_\odot$, this is the case, since the Jeans mass decreases
gradually as fragmentation proceeds. When galaxy sized masses
collapse, leading to more violent instabilities, large fluctuations,
and shocks at the origin of starbursts, they are much higher than the
Jeans mass. However, when a disk forms, rotation takes over, and also
compensates for the gravity forces at large scale.
The pressure forces stabilise the smallest scales (given by the Toomre
criterion), and rotation stabilises the large scales (allowing disk
structures to survive).  Apparently, in galactic disks
star formation occurs substantially when the stability conditions are
violated at intermediate scales. Note that the Toomre 
criterion does not prevent the formation of substructures
when the medium is rapidly cooling. We expect that the smallest
fragments are in quasi-thermal equilibrium with the microwave
background, and statistical dynamical equilibrium between coalescence
and fragmentation over the
whole hierarchy of structures (Pfenniger \& Combes 1994).  The
dissipation is thus minimised, since the temperature of the fragments is
very close to that of the cosmic background.  Clumpuscules are thought
as dynamical entities, they are frequently reshuffled
by collisional disruption and coalescence.  They permanently reform
through Jeans instability and fragmentation, so the dynamical
quasi-equilibrium is statistical along the whole hierarchy of clumps.

At high redshift ($z> 50-100$) therefore, only sporadic star-formation
(or even MACHOS formation) should occur in rare places, and that might
be sufficient to initiate the reionization and the re-heating of the
intergalactic medium (IGM), as is necessary to explain the high
ionization fraction of the Lyman-$\alpha$ absorbers (Tegmark et al.\
1984) and the omnipresent HeII gas at $z\sim 2$ (Jacobsen et al.\ 1994;
Davidsen et al.\ 1996; Reimers et al.\ 1997). The first generation of stars
could form today a small percentage of the dark halos around galaxies
under the form of white or brown dwarfs (Alcock et al.\ 1997).

\subsection{Clumpuscules resistance to reionization}
Once the bulk of the gas mass has condensed into cold molecular
hydrogen at $z\sim 150$, the clumps survive the reionization
phase of the IGM in large part, because of their high column density
($> 10^{25}$ cm$^{-2}$), that self-shields them from the radiation,
and their fractal structure, which is more gregarious than an
homogeneous distribution
in space (cf.\ Combes \& Pfenniger 1997).  Some erosion occurs, mainly
at the interfaces, which produces first atomic hydrogen, then ionized
gas. But in the densest environments, that will become the galaxies,
the bulk of the gas remains in cold clumpuscules.  The estimation of
the column density threshold of atomic hydrogen at the ionizing limit
can be obtained from the measured and modeled values of the
extra-galactic ionizing radiation, that is known now within a factor of
a few.  The extra-galactic background, essentially from the quasar UV
light, provides an ionization rate of $\xi \sim 2 \cdot
10^{-14}$\,s$^{-1}$, a value corresponding to the study of
low-redshift Lyman-$\alpha$ absorption lines (Madau 1992).  As in the
``Str\"omgren sphere theory", there is a sharp transition (a true
ionizing front) between the neutral and ionized phases, and the
transition limit is around $N_{\rm HI} \sim 10^{18}$\,cm$^{-2}$, for
usual disk gas volumic densities $n$ (it decreases as $1/n$).  This
corresponds well to the observed sudden decrease from a column density
of $N_{\rm HI} = 2 \cdot 10^{19}$\,cm$^{-2}$ to 2 $\cdot
10^{18}$\,cm$^{-2}$ in the outskirts of HI disks (Corbelli \& Salpeter
1993). Although the extra-galactic flux might increase with redshift,
it seems therefore quite easy to conserve the cold H$_2$ clumps within
their HI interface at high redshifts.

\section{H$_2$ cooling}
The details of H$_2$ chemistry and cooling has been widely
developed (Lepp \& Shull 1984; Haiman et al.\ 1996a; Tegmark et al.\ 1997).
Numerical 3D simulations have been performed (Abel et al.\ 1997; 
Anninos et al.\ 1997), in the aim of determining
the mass of the first luminous objects to form, and of following
the effective Jeans mass as a function of redshift. It was shown that
very small masses ($<0.1$\,M$_\odot$) could be formed before
reionization ($z>50-100$), but then the first stellar formation would
produce enough UV radiation to destroy the molecules (Haiman et al.\
1997), although the exact processes are complex (the UV irradiation
could even trigger extra-cooling by favoring H$_2$ formation, Haiman
et al.\ 1996b).  This H$_2$ destruction through the reionization period
is used to delay the formation of dwarf galaxies, since giant
molecular clouds will not form again until low redshifts, when
efficient cooling is provided by the metals (Norman \& Spaans 1997;
Kepner et al.\ 1997).

\begin{figure}
%\vspace{-13mm}
\centerline{\hspace{2mm}
\epsfysize=7.3cm\epsfbox{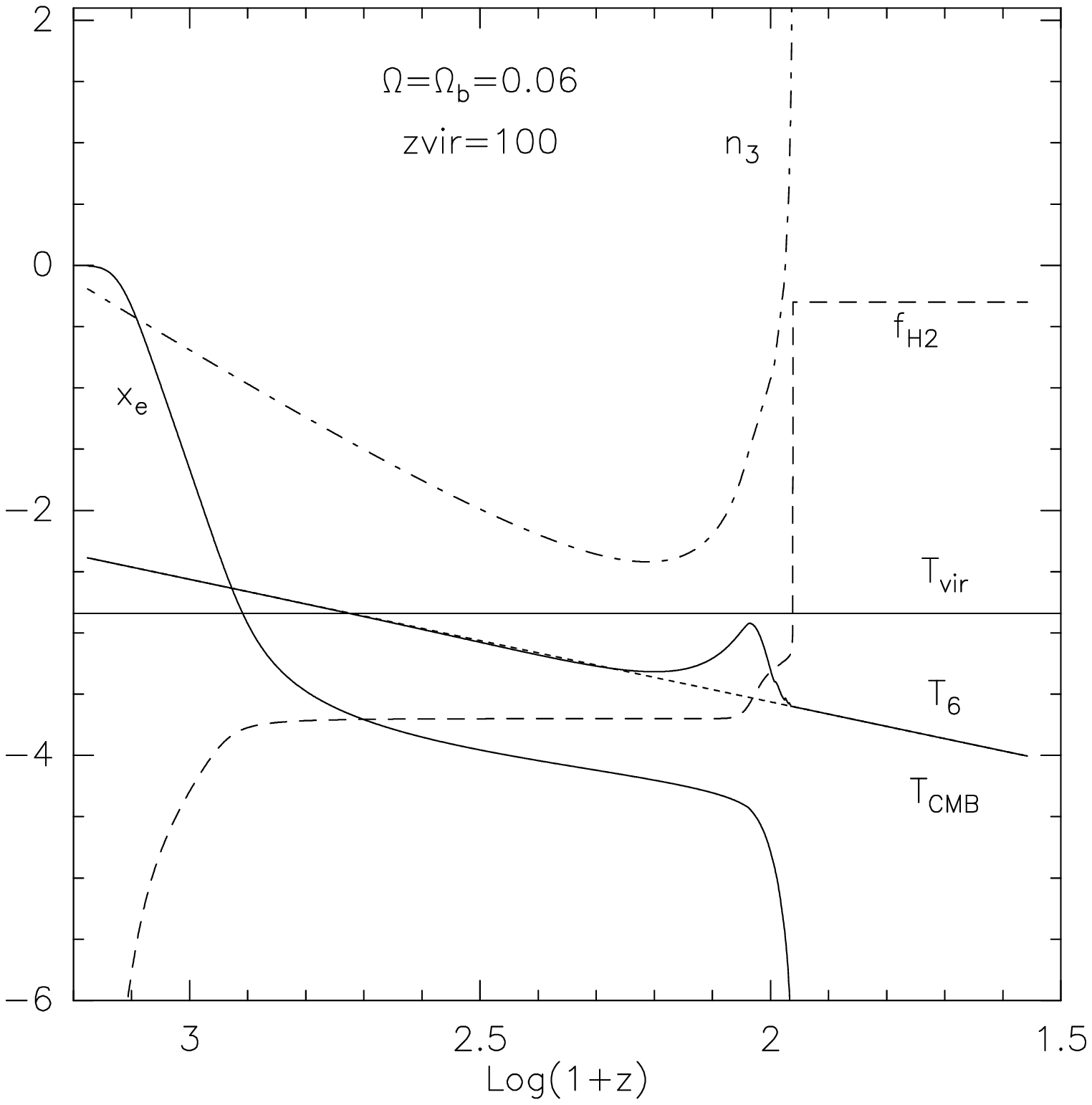}\hspace{-2mm}
\epsfysize=7.3cm\epsfbox{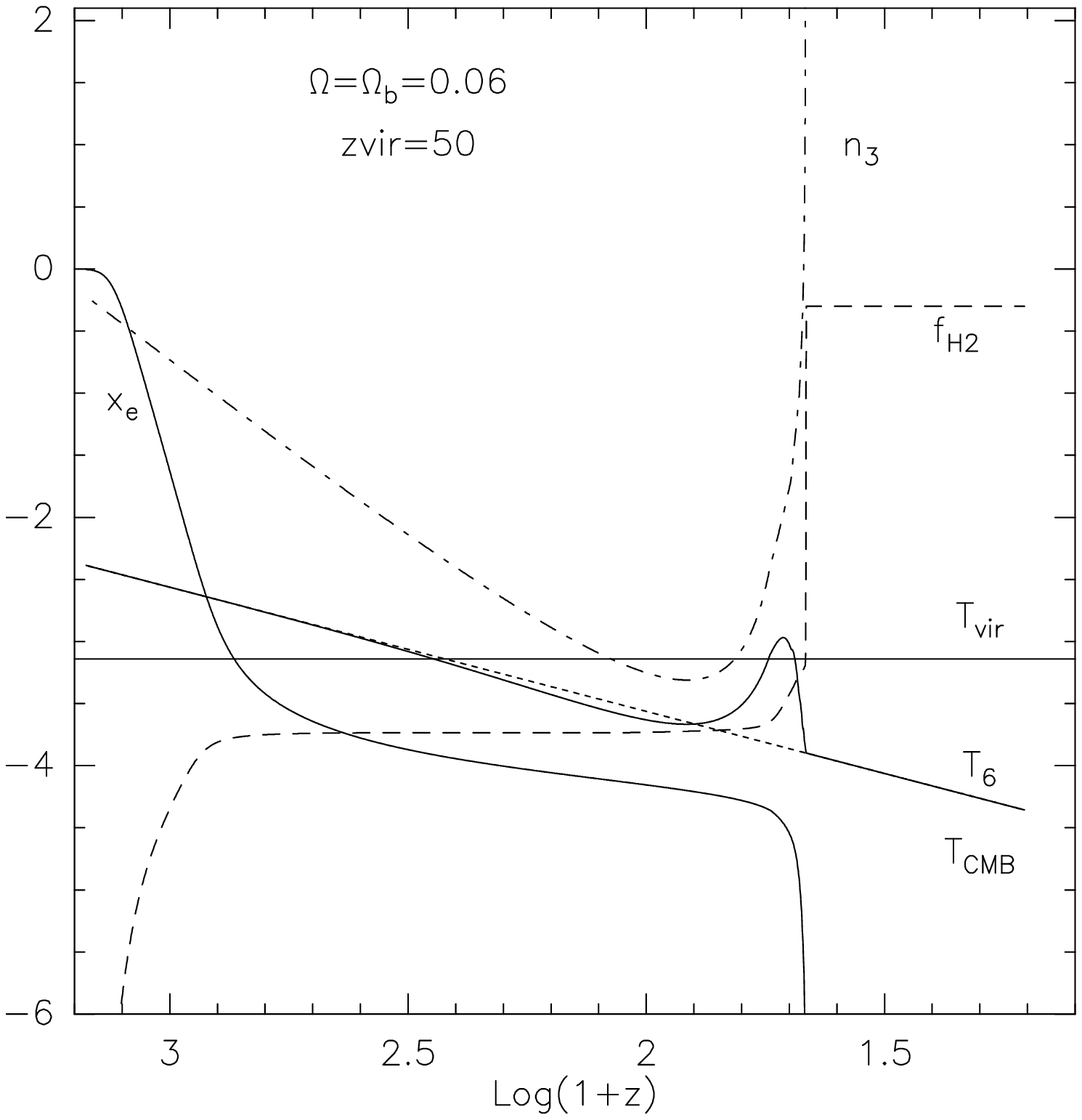}\hspace{1.8mm}}
%\vspace{-5mm}
\caption[h]{ Evolution of temperature ($T_6$ in 10$^6$\,K), molecular
fraction ($f_{\rm H_2}$), ionization fraction ($x_{\rm e}$) and
density ($n_3$ in 10$^3$\,cm$^{-3}$) for a gas cloud after
recombination, collapsing at $z_{\rm vir} = 100$ (left), and $z_{\rm
vir} = 50$ (right). The Hubble constant is $h=0.5$, and
$\Omega=\Omega_{\rm b}=0.06$. }
\end{figure}

The formation of small molecular clouds before reionization is
therefore not in doubt, and we show in a simple model resolving the
chemistry and cooling equations, that high enough densities are indeed
achieved. For the sake of comparison, we adopt
a simple scheme very similar to that of Tegmark et al.\ (1997):
since we stop when the clumpuscules become optically thick with
respect to the H$_2$ rotational lines (i.e., at column densities of
the order of $10^{24-25}$\,cm$^{-2}$, and volumic density of $\sim
10^{10}$\,cm$^{-3}$), we have to consider the gas thermal history of a
unit volume only, irrespective of the mass of the condensed
object. The density before virialisation is determined by the simple
top-hat model, and when virialised, we consider its possible recursive
fragmentation. To fix ideas, we adopt physical parameters compatible
with the observed fractal of the Milky Way molecular ISM, a fractal
dimension of $D=1.7$, and a number of fragments of $N=8$ at each level
(Pfenniger \& Combes 1994).  The ratio between the different scales of
the hierarchy is then $N^{1/D} =3.4$ and the density ratio is $r_d =
N^{(3-D)/D} =4.9$.  Assuming that the cooling is always efficient
enough to evacuate the gravitational energy of the collapse, an
hypothesis that will be confirmed a posteriori according to the virial
masses considered, we can compute the density as a function of time by
steps corresponding to the free-fall time $\tau_{{\rm ff}i}$ of a
given level $i$ of the hierarchy: $n(t) = n_i r_d^{(t-t_i)/\tau_{{\rm
ff}i}}$.  The H$_2$ molecules are formed essentially through the H$^+
+ $H (via H$_2^+$) reaction at high redshift ($z\sim 1000$, Shapiro \&
Kang 1987), then through the H$^- +$H reaction at $z\sim 100$
(Hutchins 1976), and through the 3-body reactions (H$+$H$+$H) at very
high densities (Palla et al.\ 1983). The cooling at $T< 1000$\,K is
essentially due to H$_2$.  At low temperature ($T< 100$\,K), the HD
molecules can also intervene.  The Li\,H molecules are formed in too
low quantities to be helpful (Stancil et al.\ 1996). In any case, this
occurs at so high column densities that the H$_2$ rotational lines are
optically thick, which is not considered here.  

Some results of the computations are displayed in Fig.\ 1, for $z_{\rm
vir}=100$ and 50.  The horizontal full line indicates the virial
temperature of the Jeans mass at recombination, i.e.,
$M=10^5$\,M$_\odot$. At $z_{\rm vir} = 100$, the computed gas
temperature is always lower than $T_{\rm vir}$, and therefore the
hypothesis of fragmentation is verified for all masses able to
collapse. Computations as a function of $z_{\rm vir}$ indicate that
this is true for $z_{\rm vir} > 60$. For lower $z_{\rm vir}$, only
larger masses would fragment down the whole hierarchy, unless some
metal enrichment has occurred through sporadic star formation before.
{\em Therefore we have shown that a different hypothesis about the
outcome of fragmentation leads to completely different results from
Tegmark et al.\ (1997) concerning the amount of molecular hydrogen and
the state of the bulk of the gas at redshifts $<50$}.

\section{The self-gravitating fractal gas}
\subsection{Merging tendency}
Before significant star formation has occurred in the cold gas
component, the medium is still dissipative and self-gravitating, and
ideas on quasi-isothermal fragmentation can be applied (de Vega et
al.\ 1996).  The stability against sub-clumps merging ensues.  This is
opposite to the behaviour of dissipationless components, such as
MACHOS, stars or non-baryonic dark matter. In the latter case, it is
assumed that as soon as a structure decouples from expansion, all the
sub-structures already decoupled are washed out in a dynamical time
through merging. This is due to violent relaxation and dynamical
friction, that transfer the energy of relative orbital motions to the
sub-structures themselves, and disrupt them (e.g., White \& Rees
1978). On the contrary, in a dissipative collapse and fragmentation
process, as soon as sub-structures coalesce, they form an entity
larger than the Jeans mass, and can re-fragment in a free-fall time,
or dynamical time for this scale (e.g., Pfenniger \& Combes 1994). The
fragmentation along the whole hierarchy of the Jeans mass after
recombination is almost instantaneous, as soon as the Jeans mass has
collapsed, as already noted by Hoyle (1953).  The whole process takes
only a small fraction of the collapse time-scale of the GMC at this
epoch of $\tau \sim$ 1 Myr.

\subsection{Galaxy formation}
After recombination and re-heating, the formation of larger and larger
objects is possible as long as the corresponding fluctuations become
non-linear. The formation of galaxies could range from redshifts
$z=50-100$ to the present according to their mass (at redshift $z=200$
normal 10\,kpc-radius galaxies would overlap, cf.\ Peebles
1980). Since the proto-galaxies contained mainly gas at the beginning,
dissipation forms disks, and rotation support is provided by
the angular momentum gained in tidal interactions. The proto-galaxy
collapse is therefore a rather 
violent relaxation process, and the rate of star formation could be
high (and higher in the largest mass objects).  In any case, the
relaxation is much shorter and violent in the center of the collapsed
objects. In the outer parts, most of the mass can remain there
in the form of self-gravitating cold gas, that has subsisted from
its formation period before the reionization.

When a substantial part of the mass has formed a stellar dissipationless
component, the galaxy population enters the phase of hierarchical merging.
According to their environment, and their tidal interactions with
companions, galaxies evolve at different speeds
toward the early-type side of the Hubble sequence 
(Pfenniger, Combes \& Martinet 1994).
When structures of the group-size turn around, further stellar
activity peaks occur.  Later, the formation of clusters trigger again
enhanced activity.

\subsection{Cluster formation}
Clusters form around redshift $z\sim 2$, according to their mass, and
to the cosmological model adopted. Many are still assembling today, as
is attested by the substructures observed.  What happens then to the
extended gas halos that still surround late-type galaxies? They are
very likely to be stripped, either through tidal interactions, or to
ram-pressure when a substantial and dense hot intergalactic
gas component has formed.  Already in the HI tracer, galaxies appear
to be stripped in the center of clusters like Virgo and Coma
(e.g.\ Cayatte et al.\ 1990).  Some of the stripped gas is heated at the
virial temperature of the cluster, and accounts for the huge amounts
of gas mass detected in X-rays (e.g.\ David et al.\ 1995).  But all the
clumpuscules are not destroyed, and a multi-phase medium could
survive in the IGM (Ferland et al.\ 1994), especially at the center of
the cluster, where the cooling time is shorter than the dynamical time
(cooling flows), and the cold clumpuscules can reform.
{\em Therefore the bulk of the baryonic mass detected in clusters in
the form of X-ray gas is probably only a lower limit to the existing
gas, and therefore to the amount of baryons.}

\section{Conclusions}
We propose that just after recombination and decoupling, gaseous
structures having the Jeans mass, of the order of a giant molecular
cloud of $10^{5-6}$\,M$_\odot$, collapse and fragment in the same time
to form cold molecular clumpuscules. The fragmentation can proceed to
masses down to Jupiter masses, since the collapse is quasi-isothermal,
due to the H$_2$ cooling. The large majority of these
pressure-supported fragments do not form stars, since they are in
statistical equilibrium between coalescence and fragmentation, in a
fractal structure, and in thermal equilibrium with the cosmic
background temperature. A very low level, sporadic, star formation is
sufficient to re-ionize the intergalactic gas at $z=50-100$. The bulk
of the clumpuscules survive the reionization, and may be
assembled later on in larger structures, when proto-galaxies form. This
scenario provides a way to avoid the ``cooling" catastrophy at large
redshifts, that will produce much more stars that are observed today
(Blanchard et al.\ 1992). It implies the existence of a large fraction
of the dark matter around galaxies under the form of cold molecular
gas (also consistent with the recent X-ray results in clusters, David
1997).

\end{document}